\documentclass[12pt]{iopart}
\usepackage{multicol}
\usepackage[left=3.0cm]{geometry}
\usepackage[utf8]{inputenc}
\usepackage{graphicx}
\usepackage{alltt}
\usepackage{url}
\usepackage{textcomp}
\usepackage{booktabs}
\usepackage{iopams}
\usepackage[T1]{fontenc}
\usepackage{gensymb}
\usepackage{csquotes}

\usepackage{changes}
\usepackage{cancel}
\def\al {\alpha}

\def\ba{\begin{eqnarray}}
\def\bam{\begin{array}}

\def\be{\begin{equation}}

\def\bi{\bibitem}

\def\dif{\mathrm{d}}

\def\ea{\end{eqnarray}} 
\def\ee{\end{equation}}

\def\ep{\epsilon}

\def\fr{\frac}

\def\ha{\frac{1}{2}~}

\def\LLL{\left[}

\def\RRR {\right]}

\def\si{\sigma}

\def\sim{\simeq}

\def\te{\theta}

\def\ts{\textstyle}

\def\tt{\tilde t}

\def\vf{\varphi }

\def\1{{\it one}}

\def\2{{\ts{\ha}\!}}

\def\3 {\ts{\frac{1}{3}\!}}
\def\4{\ts{\fr{1}{4}\!}}

\setdeletedmarkup{\protect{\color{red!85!black}\sout{#1}}}

\begin{document}

\title[The rotation of inertial frames by angular momentum: comparing material shells and gravitational waves]{Rotation of inertial frames by angular momentum of matter and of waves}

\author{W Barker$^\dag$, T Ledvinka$^\ddag$, D Lynden-Bell$^*$ and J Bi\v{c}\'{a}k$^\ddag$}

\address{$^\dag$ Cavendish Laboratory, Cambridge University, JJ Thomson Avenue, Cambridge, CB3 0HE, UK}
\address{$^\ddag$ Institute of Theoretical Physics, Faculty of Mathematics and Physics, Charles University,
V Hole\v{s}ovi\v{c}k\'{a}ch 2, 180 00 Prague 8, Czech Republic}
\address{$^*$ Institute of Astronomy, Cambridge University, Madingley Road, Cambridge, CB3 0HA, UK}

\begin{abstract}
We elucidate the dynamics of a thin spherical material shell with a
tangential pressure, using a new approach. This is both simpler than
the traditional method of extrinsic curvature junction conditions (which we also
employ), and suggests an expression for a `gravitational potential energy' of the shell. Such a shell, if slowly spinning, can rotationally drag the inertial frames within it through a finite angle as it collapses and rebounds from a minimum radius. Rebounding `spherical' and cylindrical pulses of rotating gravitational waves  were studied previously. Here we calculate their angular momentum and show that their rotational frame dragging is in agreement with that of the rotating spherical shell and a rotating cylindrical dust shell. This shows that Machian effects  occur equally for material and analogous `immaterial' sources.
\end{abstract}

\pacs{1315, 9440T}
\submitto{\CQG}

\section{Introduction}
H. Pfister gave a
brief summary of the history of the gravitomagnetism in \cite{Pfister2014}. In 1995 I. Ciufolini and J. A. Wheeler published the monograph, \cite{CiufoliniWheeler1995}, on various aspects of gravitomagnetism and on the origin of inertia within general relativity.

Various interpretations of Mach's Principle were summarized comprehensively in an excellent book, \cite{BarbourPfister1995}, which contains talks and includes also discussions from the 1993 Tuebingen conference.

The statement of Mach's Principle which we use is that the local inertial frame at any event is determined by some average of the distribution of energy and momentum. We demonstrated how this formulation acquires concrete mathematical content in our study of linear perturbation of closed universes, \cite{1995MNRAS.272..150L}, and of Friedmann-Robertson-Walker universes with any curvature and cosmological constant, \cite{bklb2007}.

However, the first-order perturbation theory could not reveal the influence of gravitational waves. The question of whether and how some (averaged) stress tensor of gravitational waves influences local inertial frames remained unclear for long time. Going back to the origins, Einstein wrote in 1917, \cite{Lorentz}, \cite{1917SPAW.......142E}, when he already knew about gravitational waves: "In a consistent theory of relativity there can be no inertia relatively to `space', but only an inertia of masses relatively to one another...". The idea that the metric tensor and thus inertia in a Machian solution of general relativity must be determined completely by the distribution of the energy-momentum tensor of matter, with no contribution from gravitational waves was realized in the late 60s in the independent works by Al'tshuler, Lynden-Bell, Sciama, Waylen and Gilman, and then developed further by Raine (see his review, \cite{raine}).

   We studied the effects of gravitational waves on local inertial frames in \cite{BLK2008a}, \cite{cylinders} and \cite{spheres}. We investigated the effects of the waves in the second order perturbation schemes on Minkowski background rather than in a more complicated cosmological background. First we studied the rotation of the inertial frame in an approximately flat cylindrical region surrounded by an ingoing and then outgoing pulse of gravitational waves rotating about the axis of cylindrical coordinates in \cite{cylinders} and \cite{BLK2008a}. The effect of angular momentum of the waves on the rotation of the inertial frames demonstrates explicitly that any formulation of Mach's principle must necesserily involve the influence of waves, not
only of material $T_{jk}$. This work was generalized to the case of a time-symmetric ingoing and then outgoing regular pulse of rotating gravitational waves propagating in empty asymptotically flat spacetime, \cite{spheres}. A nonvanishing angular momentum of the waves keeps them away from the origin where the spacetime is approximately flat. By solving the relevant Einstein equations to second order in the amplitude of the waves,
it is seen that the rotation of local inertial frames near the origin is without time delay and follows from the constraint equation.

The relevance of the present paper to Mach's Principle is that it shows explicitly that the angular momentum of gravitational waves has the same effect on the inertial frame as the angular momentum of a spherical material shell bouncing due to a tangential pressure. We find strong and quantitative similarities between these cases.

In order to see this in depth we here give the results of the calculations of the angular momentum and the effective energy of the rotating waves in asymptotically flat spacetime. This was not done in our previous work, \cite{spheres}. Although the `starting expressions' are indeed quite complicated, the final results turn into simple, intuitive forms allowing detailed comparison between rebounding material shells and gravitational waves.

We also present videos demonstrating the evolution of the rotating bouncing shells and waves, and their effect on the rotation of an inertial frame at the centre where the spacetime is very nearly flat.

The weak gravomagnetic effects of rotating cylindrical and spherical material shells might be found by, for example, constructing them from non-relativistic dust and applying the gravitational Biot-Savart law. For a cylinder of angular momentum per unit length $l_{z}$ and radius $P$, and a sphere of angular momentum $L_{z}$ and radius $R$, the uniform gravomagnetic fields inside have strengths $B_{g}=8l_{z}/P^{2}$ and $B_{g}=4L_{z}/R^{3}$ respectively. By implementing the gravitational Larmor theorem we may transform away these fields, finding ourselves in inertial frames rotating at
\begin{equation} \label{cylindrical}
	\omega_{(\mathrm{C_m})}=\frac{4l_{z}}{P^{2}}
\end{equation}
within the cylinder and
\begin{equation} \label{spherical}
	\omega_{(\mathrm{S_m})}=\frac{2L_{z}}{R^{3}}
\end{equation}
within the sphere, with respect to the observer at infinity. Curiously, the dragging at the origin from a cylinder is the same as that of the radially inscribed sphere $R=P$ if the $L_{z}$ of the sphere is replaced by the angular momentum of that part of the cylinder which lies within $45\degree$ of $z=0$, i.e. the part of the cylinder which also vertically inscribes the sphere. Analogous effects are found with spinning, charged spheres and solenoids in classical electromagnetism.

In \sref{2} we analyse the motion of a rebounding shell using a new method simpler than Israel's junction-matching procedure, \cite{shells}, (as employed by e.g. Evans, \cite{shame}). We consider the rotation of the flat internal space caused by a perturbative rotation of the shell. In \sref{3} we consider the rotation of inertial frames at the centre of the rotating, rebounding cylindrical wave pulse introduced in \cite{cylinders}. Finally in \sref{4} we inspect the same effects as caused by the rotating, rebounding `spherical' wave pulse introduced in \cite{spheres}. Conclusions follow.

Latin and Greek indices are understood to run over spaces of signature $-2$ and $-1$ respectively and we use the Planck units, $c=G=1$.

\section{Rotating and rebounding material shell} \label{2}
\subsection{Equation of motion} \label{3.1}
Consider a heavy thin spherical shell made up from counter-rotating collisionless equal particles all of which have the same magnitude of angular momentum. We are interested in the radial motion of the whole shell. Let the total rest mass of all the particles be $M$  and the gravitational mass of the external Schwarzschild metric be $m_g$; we define the ratio $a=m_g/M$ as in \cite{causality}.
At any one moment we may consider the whole shell to be made up of nested sub-shells living in the Schwarzschild metric,
\be
\dif s^2=(1-2\mu/r)\dif t^2-(1-2\mu/r)^{-1} \dif r^2-r^2(\dif\te^2+\sin^2\te \dif\vf^2), 
\ee
where the parameter $\mu$ varies from $0$ for the innermost subshell to $m_g$ for the outermost. The specific energy of particles in the subshell $\mu$ is given by $\ep(\mu)=(1-2\mu/r)dt/d\tau_p$
where $\tau_p$ is the proper time measured on the particle, and the specific angular momentum of motion (which without loss of generality we can take to be equatorial) is $h=r^2\dif\vf/\dif\tau_p$ so its motion is governed by
\be
1=(1-2\mu/r)^{-1} \left[\ep^2-(\dif r/\dif\tau_p)^2\right]-h^2/r^2.
\ee
Proper time on the shell at constant 
$\te$, $\vf $ is given by $\dif\tau^2=(1+h^2/r^2)\dif\tau_p^2$. For a thin shell, all subshells share the same radial coordinate, $R$, and writing $\dot{R}=\dif R/\dif\tau$ the specific energy is
\be
\ep(\mu) =\sqrt{(1+h^2/R^2)(1-2\mu/R+\dot{R}^2)}.
\ee
The subshell that contributes $\dif\mu$ to $m_g$ contributes only $\dif\mu/\ep(\mu)$ to $M$ so that 
\be
\hspace*{-2cm} 
M=\int _{0}^{m_g} \frac{\dif\mu}{\sqrt{(1+h^2/R^2)(1-2\mu/R+\dot{R}^2)}}=\LLL(-R)\sqrt{\frac{1-2\mu/r+\dot{R}^2}{1+h^2/R^2}}~~\RRR^{m_g}_{0},
\ee
or
\be\label{eqmotionp}
\sqrt{1+h^2/R^2}~M/r-\sqrt{1+\dot{R}^2}=-\sqrt{1-2m_g/R+\dot{R}^2}.
\ee
Squaring and solving for $m_g$ which is the total energy of the system gives
\be\label{eqmotion}
m_g=M\sqrt{(1+h^2/R^2)(1+\dot{R}^2)}-\2 ~(1+h^2/R^2)M^2/R.
\ee
By considering the contribution of each moving particle, the first term is seen to be equivalent to the total kinetic energy of the shell, the second to the gravitational potential energy.
\par
If $h$ vanishes, we recover the well-known result for dusty shells derived by Israel in \cite{shells}:
\begin{eqnarray}
\sqrt{1+\dot{R}^{2}}=a+\frac{m_{g}}{2aR}.
\end{eqnarray}
\par
A more conventional derivation of \eref{eqmotion} using the method of Israel is presented in \ref{A}.

\subsection{Dynamics of the shell} \label{3.2}
\begin{figure}[h]
\centering
\includegraphics[width=12cm]{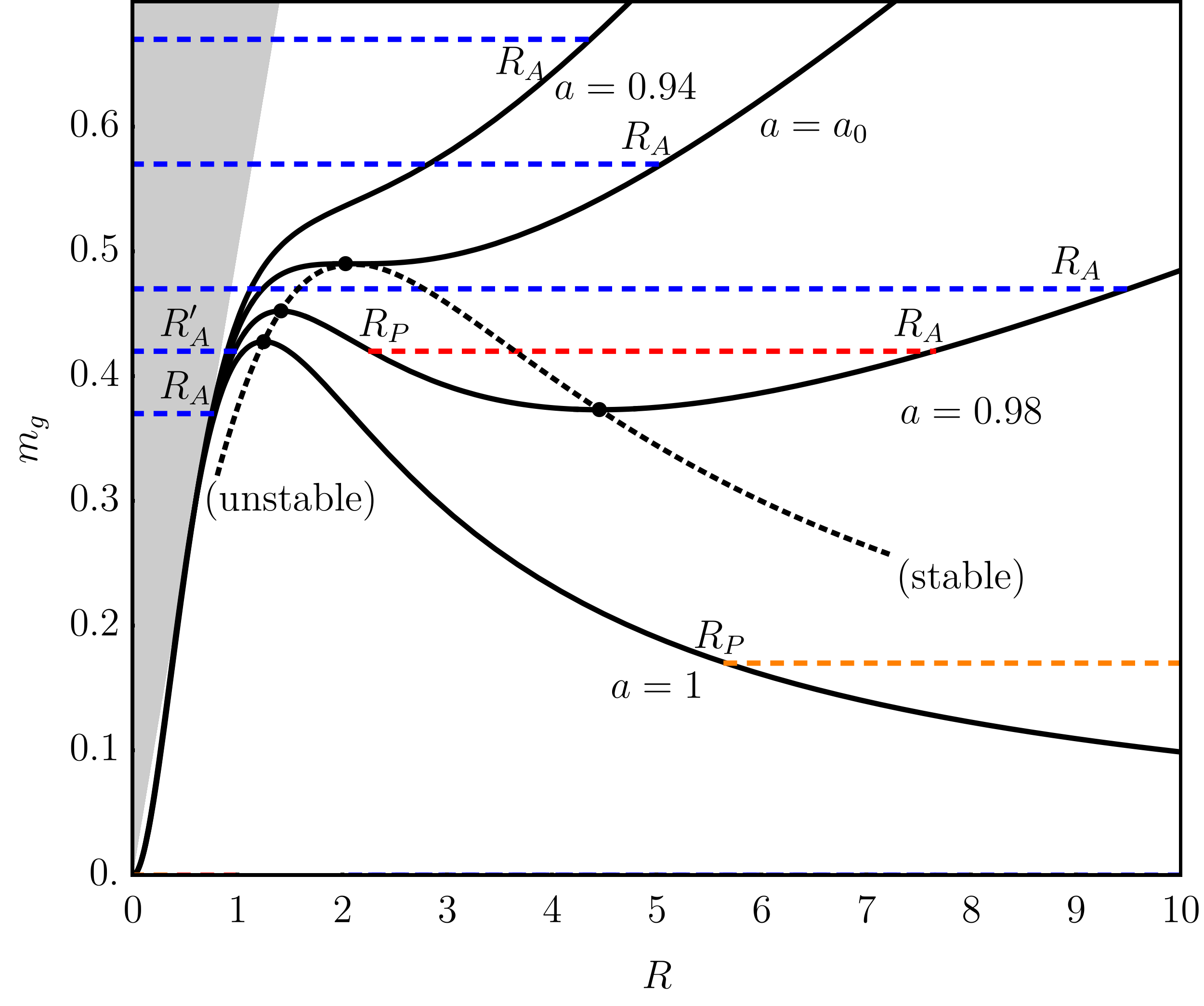}
\caption{$m_g=hf(a,h/R)$ as a function of $R$ for various $a$ at $h=1$ with space below the gravitational radius, $r_g$, shaded in grey. A rebounding collapse from rest at infinity is shown in orange, the oscillating trajectory in red and collapse to a black hole from rest at finite radius in blue. Israel's principle of submerged shells applies because the curves only ever brush $r_g$, and the only effect of specific angular momentum of the particles, $h$, is to dilate the curves in $m_g-R$ space.}
\label{fig_2}
\end{figure}

To categorise the various trajectories of the shell, we determine the stationary points of the radial motion, i.e. the zeroes of $\dot{R}$. We introduce the dimensionless quantity $\xi=h/R$
and use \eref{eqmotion} to express its dependence on the parameters $a$, $h$ and $m_{g}$ as an implicit function:
\begin{eqnarray} \label{eq:teller}
m_{g}=hf\left(a, \xi\right)=\frac{2ah\left(\sqrt{1+\xi^{2}}-a\right)}{\xi\left(1+\xi^{2}\right)}.
\end{eqnarray}
Through variation of $a$, $m_{g}$ and $h$ two radial turning points may merge or bifurcate at a point in the parameter space corresponding to circular orbits of the particles in the shell. The condition for this is $\partial f\left(a, \xi\right)/\partial \xi=0$, which can eventually be written as
\begin{eqnarray} \label{eq:chi}
g\left(a, \xi\right)=4\xi^6+\left(8-9a^{2}\right)\xi^4+\left(5-6a^{2}\right)\xi^{2}+1-a^{2}=0.
\end{eqnarray}
At $a=0$, \eref{eq:chi} has no solution since all the coefficients are positive. As $a$ is increased from $0$ two zeros of $g\left(a, \xi\right)$, $\xi=\xi_{1}$ and $\xi=\xi_{2}$, bifurcate from a double zero, $\xi=\xi_{0}\doteq 0.478$ at $a=a_0\doteq 0.958$. As $a$ increases further from $a_0$ to $1$, $\xi_{1}$ vanishes, meanwhile $\xi_{2}$ continues to increase indefinitely with $a$. This corresponds to the development of a local minimum and maximum in $f\left(a, \xi\right)$ from an inflection point at $a_0$. The dimensionless number $a_0$ is a fundamental constant of this particular class of shells, it is given by the simultaneous solution of
\eref{eq:chi} with the condition $\partial g\left(a, \xi\right)/\partial \xi=0 $. As shown in \Fref{fig_2} the dependence of the trajectories of the shell on $a$ can be broken into several contingencies.
\par
When $a=1$ the gravitational mass coincides with the sum of the particle rest masses and the only circular orbit is at $\xi_{2}=\frac{1}{2}\sqrt{1/2+\sqrt{17}/2}\doteq0.80$. By expanding the effective potential about this point it is clear that it represents a merger of two \textit{distinct} shell trajectories, so the circular orbit is completely unstable. These trajectories are a collapse from rest at infinity followed by a rebound at a pericenter, $R_{P}$, and a collapse to a black hole either from rest at an apocenter, $R_{A}<R_{P}$, or from rest at infinity. $\dot{R}$ diverges as $R^{-2}$ on the approach to singularity. 
\par
When $a>1$, the shell's kinetic energy exceeds its gravitational potential energy, so the shell is always in motion at infinite radius. 
\par
Alternatively, if $a<1$ but $a_0<a$, the circular orbit at $\xi_{1}$, which is completely stable, must also be taken into account. As $a$ drops below $1$ the asymptote in $f\left(a, \xi\right)$ as $R\rightarrow\infty$ disappears so the shell never reaches an infinite radius, instead falling from rest at a new apocenter, $R_{A}'>R_{A}$. This allows for a new trajectory where the shell oscillates between the two radii $R_{A}'$ and $R_{P}$.
\par
Finally for $a<a_0$, $R_A=R'_A$ and the oscillating trajectory vanishes. Thus $a_0$ is interpreted to be the minimum ratio of gravitational to nucleonic mass for which the shell can resist collapse to a black hole. The circular orbit $\xi_{0}$ is stable against expansion but unstable to collapse.
\paragraph{}
It is neatly found in \cite{causality} that "no shell of dust can remain permanently submerged within its Schwarzschild sphere". By solving \eref{eq:teller} for $a$ we find the parameters of the shell which collapses from $r_{g}$,
\begin{eqnarray} 									          
a=\frac{1}{2}\sqrt{1+h^{2}/4m_{g}^{2}},
\end{eqnarray} 
which reduces to $a=1/2$ for dusty shells as found in \cite{causality}. Since this solution is unique for nonzero physical parameters, and since $\partial f\left(a, h/R\right)/\partial R$ vanishes at the origin and approaches $2a\left(1-a\right)$ for large $R$, it is clear that $R_{A}$ never drops below $r_g$ and so this principle is preserved, independent of $h$. There are no shells which collapse from $r_g$ for $a<1/2$.

\subsection{Pertubative rotation and conservation of angular momentum}	\label{3.3}
We now spin the shell about its polar axis with total angular momentum $L_{z}$, using the formalism developed in \cite{perturbed} for the case of slowly rotating, collapsing dusty shells. Working to $\mathcal{O}\left(r\omega\right)$ the exterior space now has a Kerr-like perturbation,
\be
\hspace*{-2cm} 
\mathrm{d}s^{2}=\left(1-r_{g}/r\right)\mathrm{d}t^{2}-\left(1-r_{g}/r\right)^{-1}\mathrm{d}r^{2}-r^{2}\left[\mathrm{d}\theta^{2}+\sin^{2}\theta\left(\mathrm{d}\varphi^{2}-2\omega \mathrm{d}\varphi \mathrm{d}t\right)\right],
\ee
where $\omega=2L_{z}/r^{3}$ or at the shell itself $\omega_{(\mathrm{S_m})}=2L_{z}/R^{3}$. The interior space is flat,
\begin{eqnarray}
\mathrm{d}s^{2}=\mathrm{d}\bar{t}^{2}-\mathrm{d}\bar{r}^{2}-\bar{r}^{2}\left(\mathrm{d}\theta^{2}+\sin^{2}\theta\mathrm{d}\bar{\varphi}^{2}\right),
\end{eqnarray}
but the azimuthal coordinates $\vf$ and $\bar{\vf}$ no longer generally match at the shell itself, which can have the coordinates,
\begin{eqnarray}
\mathrm{d}s^{2}=\mathrm{d}\tau^{2}-R^{2}\left(\mathrm{d}\theta^{2}+\sin^{2}\theta\mathrm{d}\bar{\varphi}^{2}\right).
\end{eqnarray}
It can be shown (see \ref{B}) that the sphericity and equation of motion of fluid shells are unchanged by the rotation to $\mathcal{O}\left(R\omega_{(\mathrm{S_m})}\right)$. An important relation used in \cite{perturbed} (see \ref{C}) is
\begin{eqnarray} \label{rota}
\tau^{0}_{3}=-\frac{1}{16\pi }\frac{\mathrm{d}\omega_{(\mathrm{S_m})}}{\mathrm{d}\frac{1}{R}}\sin^{2}\theta,
\end{eqnarray}
where $\tau^{\alpha}_{\beta}$ is the stress-energy tensor in the shell hypersurface. If the rotation angle in the shell hypersurface (measured in $\bar{\varphi}$) of the fluid of the shell as it rotates is denoted by $\bar{\Phi}$, and analogously in the exterior (measured in $\vf$) by $\Phi$, there is to $\mathcal{O}\left(R\omega_{(\mathrm{S_m})}\right)$,
\begin{eqnarray}
\tau^{0}_{3}=-\frac{m_{g}}{4\pi  aR}\left(\sqrt{R^{2}+h^{2}}+\frac{h^{2}}{2\sqrt{R^{2}+h^{2}}}\right)\dot{\bar{\Phi}}\sin^{2}\theta,
\end{eqnarray}
or
\begin{eqnarray} \label{eq:relat}
\dot{\bar{\Phi}}=\frac{3aR^{2}\omega_{(\mathrm{S_m})}}{4m_{g}}\left(\sqrt{R^{2}+h^{2}}+\frac{h^{2}}{2\sqrt{R^{2}+h^{2}}}\right)^{-1}.
\end{eqnarray}
If $h$ vanishes, we recover the analogous result for dusty shells, found in \cite{perturbed}:
\begin{eqnarray} \label{eq:relatdust}
\dot{\bar{\Phi}}=\frac{3aR\omega_{(\mathrm{S_m})}}{4m_{g}}.
\end{eqnarray}
\par
It is useful to note that it is possible to write down $L_{z}$ for any rotating spherical shell immediately by transforming the unperturbed $\tau^{\alpha}_{\beta}$ with the rotation to find momentum density as seen in the shell hypersurface, then integrating over the angular coordinates. For perfect fluid spheres and to $\mathcal{O}\left(R\omega_{(\mathrm{S_m})}\right)$,
\begin{eqnarray}
L_{z}=2\pi\int^{\pi}_{0}\mathrm{d}\theta\dot{\bar{\Phi}} R^{4}\sin^{3}\theta\left(\sigma+\Pi\right),
\end{eqnarray}
where $\si=\tau^{0}_{0}$ is the mass-energy density and $\Pi$ is the tangential pressure (see \ref{A}). This confirms conservation of $L_{z}$ as it rotates according to \eref{eq:relat}.
\par
The evolving azimuthal discontinuity at the junction is the direct manifestation of the rotational frame dragging of the interior. It is pointed out in \cite{perturbed} that
\begin{eqnarray}
\frac{\mathrm{d}\bar{\varphi}}{\mathrm{d}t}=0 \quad \Longrightarrow \quad \frac{\mathrm{d}\varphi}{\mathrm{d}t}=\omega_{(\mathrm{S_m})},
\end{eqnarray}
so that external observers falling with the shell and at angular velocity $\omega_{(\mathrm{S_m})}$ see a nonrotating interior inertial frame, equivalently observers at rest at infinity see the interior inertial frame rotating at $\omega_{(\mathrm{S_m})}$.
\par
For a spinning $a=1$ shell which falls from rest at infinity and rebounds from a pericenter (see subsection \ref{3.2}),
the consequent rotation of the inertial frame, and the time taken according to the observer at infinite radius are
\begin{eqnarray} \label{eq:delphi}
\Delta\varphi=2\int_{R_{P}}^{\infty}\frac{\omega_{(\mathrm{S_m})}\dot{T}\mathrm{d}R}{\dot{R}}, \quad \Delta t=2\int_{R_{P}}^{\infty}\frac{\dot{T}\mathrm{d}R}{\dot{R}}.
\end{eqnarray}
Both integrands usually contain an integrable singularity at the lower limit, the exception being the limit of asymptotic collapse to the unstable circular orbit of particles where the expansion of $\dot{R}$ changes from $\dot{R}\propto\sqrt{R-R_{P}}$ to $\dot{R}\propto\left(R-R_{P}\right)$. Then the integrals are over logarithmic singularities and so $\Delta\varphi$ and $\Delta t$ diverge. Conversely in the upper limit, the integrand of $\Delta t$ approaches $2a/\sqrt{a^{2}-1}$ or $2\sqrt{R/m_{g}}$ for $a>1$ and $a=1$, however $\Delta \varphi$ converges in this limit due to the factor of $2L_{z}/R^{3}$, so it is reasonable to expect finite rotations of the inertial frame in albeit long times\footnote{Rough estimates to this rotation may be given by the Laplace approximation,
$\Delta\varphi=\left(2L_{z}/R^{3}\right)\sqrt{2\pi R_{P}/3\ddot{R}\left(R_{P}\right)}$, although this too must fail near $\chi_{2}$ where $\ddot{R}$ disappears}.

To illustrate this behaviour, animations are available from \url{http://utf.mff.cuni.cz/~ledvinka/psi/a1.mp4} and \url{http://utf.mff.cuni.cz/~ledvinka/psi/a2.mp4}. The shell (outer sphere) rebounds from $R_{P}$ above its $r_{g}$ (inner sphere). The time and the angle $\Phi$ through which the fluid of the shell rotates are those of the observer at infinity, who sees the $\bar{\varphi}$ inside rotationally dragged. Blue arrows are for reference, angular momenta ($J=L_{z}$) are made artificially large and the observer at constant $\theta$, $\bar{\vf}$ whose proper time is $\tau$ is marked by the point $P$. In the first animation, $R_{P}$ is far from $r_{g}$, and during the prompt rebound the rotation of inertial frames is very slight compared to that of the shell fluid. Conversely the second animation, $R_{P}$ is close to the unstable particle orbit and whilst the shell lingers on the brink of gravitational collapse near to $r_{g}$ the rotation of the frames is much stronger.

\section{Rotating and rebounding cylinder of gravitational waves} \label{3}
In \cite{cylinders} a cylindrical pulse of gravitational waves parametrised by a timescale $A$\footnote{Where this characteristic width $A=a$ as it appears in both \cite{cylinders} and \cite{spheres}, so as to avoid confusion with the unrelated quantity $m_g/M=a$ as it appears in both \cite{causality} and in the current paper} is analysed, with relevant quantities being evaluated in the dimensionless coordinates $\tilde{\rho}=\rho/A$ and $\tilde{t}=t/A$. The first order metric perturbations are proportional to the small function,
\begin{eqnarray}
\psi\propto B\tilde{\rho}^{m}\left[\left(1+\tilde{\rho}^{2}-\tilde{t}^{2}\right)^{2}+4\tilde{t}^{2}\right]^{-\frac{1}{2}\left(m+\frac{1}{2}\right)},
\end{eqnarray}
where $B$ acts as an amplitude and we neglect a trigonometric factor in $\tilde{t}$ and $\tilde{\rho}$ which produces the spiralling wavefronts: looking down the $z$-axis, this is the radial envelope of the ring-shaped wavepacket. The maximum of this envelope is given at $\tilde{P}$ such that
\begin{eqnarray}
\frac{m}{\tilde{P}}=\frac{2\left(m+1/2\right)\left(1+\tilde{P}^{2}-\tilde{t}^{2}\right)\tilde{P}}{\left(1+\tilde{P}^{2}-\tilde{t}^{2}\right)^{2}+4\tilde{t}^{2}},
\end{eqnarray}
or
\begin{eqnarray}
\tilde{P}^{2}=\frac{\tilde{t}^{2}-1+\sqrt{\left(\tilde{t}^2-1\right)^{2}+4m\left(1+m\right)\left(\tilde{t}^{2}+1\right)^{2}}}{2\left(m+1\right)}.
\end{eqnarray}
As expected for large times the radial pulse approaches the speed of light, $\tilde{P}=\tilde{t}$, but slows as it approaches the central axis. The rate of rotational frame dragging, $\omega$ (as perceived by the distant observer), is generally calculated as an average over the azimuthal coordinate of the metric describing the waves, an operation denoted by $\left\langle\right\rangle$; in the nearly flat space at the central axis it simply reduces to the formula given in \cite{cylinders},
\begin{eqnarray}
\left\langle\omega\right\rangle\bigg|_{\tilde\rho=0}=\omega_{(\mathrm{C_w})}=\frac{B^{2}}{A}\cdot\frac{\left(2m\right)!}{2^{2m-1}}\cdot\frac{1+m\left(1+\tilde{t}^{2}\right)}{\left(1+\tilde{t}^{2}\right)^{2}}.
\end{eqnarray}
From here we may write the rate of on-axis rotational frame dragging as a function of this radius for various $m$:
\begin{eqnarray}
\omega_{(\mathrm{C_w})}=\frac{B^{2}}{A}\cdot\frac{\left(2m\right)!}{2^{2m-1}}\cdot\left(2m^{2}\right)\cdot\frac{2+\tilde{P}^{2}\left(\sqrt{\left(2m+1\right)^{2}+\frac{8m}{\tilde{P}^{2}}}-1\right)}{\tilde{P}^{4}\left(\sqrt{\left(2m+1\right)^{2}+\frac{8m}{\tilde{P}^{2}}}-1\right)^{2}}.
\end{eqnarray}
The formula for the angular momentum per unit height of the cylinder is
\begin{eqnarray}
l_{z}=\frac{A B^{2}}{2}\cdot\frac{m\left(2m\right)!}{2^{2m}}
\end{eqnarray}
and so we may eliminate the $B$ in favour of the more general quantity $l_{z}$ to give
\begin{eqnarray} \label{eq:complex}
\omega_{(\mathrm{C_w})}&=\frac{4l_{z}}{mA^{2}}\cdot\frac{1+m\left(1+\tilde{t}^{2}\right)}{\left(1+\tilde{t}^{2}\right)^{2}}\\&=\frac{8l_{z}m}{A^{2}}\cdot\frac{2+\tilde{P}^{2}\left(\sqrt{\left(2m+1\right)^{2}+\frac{8m}{\tilde{P}^{2}}}-1\right)}{\tilde{P}^{4}\left(\sqrt{\left(2m+1\right)^{2}+\frac{8m}{\tilde{P}^{2}}}-1\right)^{2}}.
\end{eqnarray}
\begin{figure}[t]
\centering
\includegraphics[width=12cm]{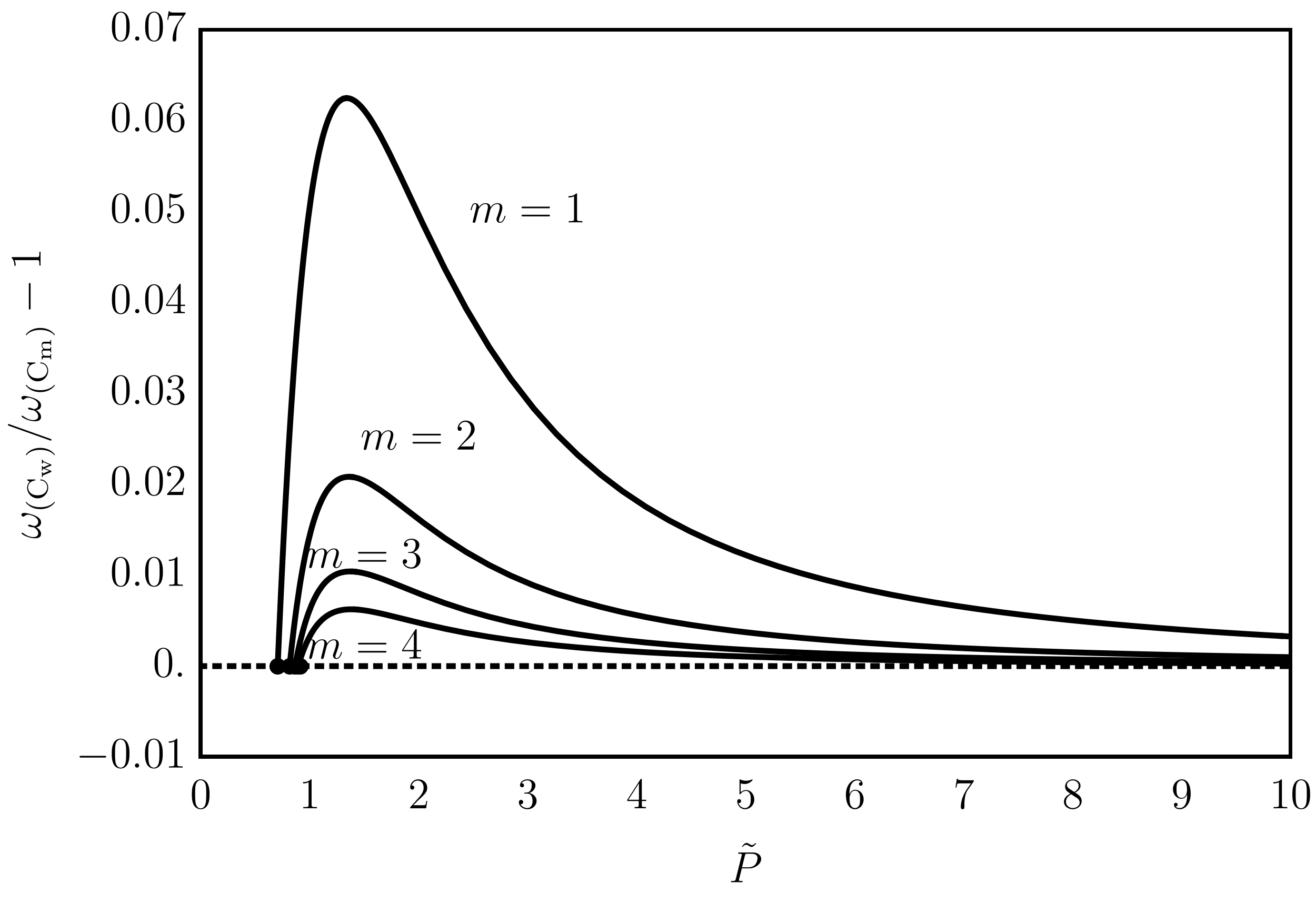}
\caption{For $l_{z}=A=1$, relative deviation of $\omega_{(\mathrm{C_w})}$ caused by wave pulses of various $m$ from the $\omega_{(\mathrm{C_m})}$ of the material cylindrical shell at $\tilde{P}$. The cylindrical wave pulse always rotates the central inertial frame faster than the material cylinder, though there is agreement at large $\tilde{P}$ and also at the minimal $\tilde{P}$, at which the waves rebound.}
\label{fig:rates}
\end{figure}
This formula for the rate of rotational frame dragging reduces to that of the material cylinder both at large absolute times where $\tilde{P}$ is also large, and at $\tilde t=0$ where the wavepacket is closest to the central axis and $\rho$ is minimal at $\tilde{P}^{2}=a^{2}m/\left(m+1\right)$. Plots of the wave-pulse in \cite{cylinders} show that the radial confinement of the wave crests increases at large  times, and so by comparison with relation \eref{cylindrical} this result is not surprising. Conversely, at $\tilde t=0$ the wavepulse is hardly recognisable as a cylinder, being highly anisotropic; so it is interesting that \eref{cylindrical} is also recovered in the instant of rebounding. This behaviour is illustrated in Figure \ref{fig:rates}.

\section{Rotating and rebounding shell of gravitational waves in asymptotically flat spacetime} \label{4}
\paragraph{}
Rotating cylindrical waves allow relatively simple comparison of their gravomagnetic effects with those from a slowly rotating collapsing spherical shell.	\par
However, in contrast to the shell their behaviour both at radial infinity and along the axis of symmetry is significantly different because there the spacetime is not asymptotically flat. Interestingly, a much more complicated case of rotating gravitational waves in an asymptotically flat vacuum spacetime allows us to make a very telling comparison between the shell and the waves.	\par
In \cite{spheres} we solve the Einstein equations to first order and form a time-symmetric ingoing and outgoing regular pulse in asymptotically flat spacetime. The waves are assumed to have odd parity and they keep away from region around the origin due to their non-vanishing angular momentum. In that region the spacetime is almost flat as it is inside a slowly rotating spherical shell. Indeed, we produced an ingoing and outgoing pulse of typical width $A$, localized in the radial direction, resembling an infalling and expanding rotating shell. We solved the relevant Einstein equation (cf. eqs. (6.8)-(6.12) in \cite{spheres}) to the second order and thus found the metric component 
$g_{t\varphi}$ in the form
\begin{eqnarray}
	g_{\ \ \ t\varphi }^{(2)}= - \omega_{(\mathrm{S_w})} \, r^2\sin^2\theta .
\end{eqnarray}
Here the angular velocity $\omega_{(\mathrm{S_w})}$ determining the rotation of an inertial frame located near the origin is given by
\begin{eqnarray}
	\omega_{(\mathrm{S_w})}=  \frac{1}{4 \pi} \int_0^\infty  \int R_{\ \ \ t\varphi }^{(2)}\left[h^{(1)} ,h^{(1)}\right]~ \mathrm{d}\Omega~ \frac{\mathrm{d}r}{r},
	\label{omega0}
\end{eqnarray}
where $\mathrm{d}\Omega = \sin \theta\, \mathrm{d}\theta\, \mathrm{d}\varphi$ and $h^{(1)}$ denotes the first order perturbations. 
All relevant $h^{(1)}$s are determined by the derivatives of the function $\chi\left(t, r,\theta,\varphi\right) = r^2 \psi_{lm}$, where $\psi$ satisfies the flat space wave equation; in contrast to cylindrical waves it depends on two spherical harmonic indexes $l$ and $m$. The second-order Ricci tensor determining, after integration, the angular velocity $\omega_{(\mathrm{S_w})}$, is given in terms of derivatives of $\chi$  by a rather lengthy formula which will not be reproduced here (see Equation (7.6) in \cite{spheres}).
\paragraph{}
To see the pulse and the rotating character of the first-order wave let us first explicitly write down the form of the `potential' function, $\chi$. It reads
\begin{eqnarray}
	\label{chi_fce}
	\chi\left(t,r,\theta,\varphi\right) &={\tilde B}_l  N^m_l {\rm Re}\left[  \frac {{r}^{l}  {P_l^m}\left(\cos\theta\right){e^{im\varphi}}}{{A}^{l+2}\left ({r}^{2}+\left (A+it\right )^{2}\right )^{l+1}} \right]\\
	&={\tilde B}_l N^m_l \kappa\left(t,r\right) P_{l}^m\left(\cos\theta\right)  \cos\left( m\varphi-\lambda\left(t,r\right) \right),
\end{eqnarray}
where, writing ${\tilde r}=r/A$ and ${\tilde t}=t/A$,
\begin{eqnarray}
	N^m_l &= \sqrt{\frac{2l+1}{4\pi} \frac{\left(l-m\right)!}{\left(l+m\right)!}}, \quad
	\kappa =\frac{\tilde r^{l+2} }{\left[ \left(1+\tilde {r}^2-\tilde {t}^2\right)^2 + 4 \tilde {t}^2 \right]^{\left(l+1\right)/2} },
\end{eqnarray}
\begin{eqnarray}
	\lambda\left(t,r\right) &= \left(l+1\right) \arctan\left(\frac{2\tilde t}{1+{\tilde r}^2-{\tilde t}^2} \right)
	=  {\rm arg}~\left[{\tilde r}^2+\left(1+i{\tilde t}\right)^2\right]^{\left(l+1\right)}.
\end{eqnarray}
\begin{figure}[ht!]
	\begin{center}
		\includegraphics[width=9cm]{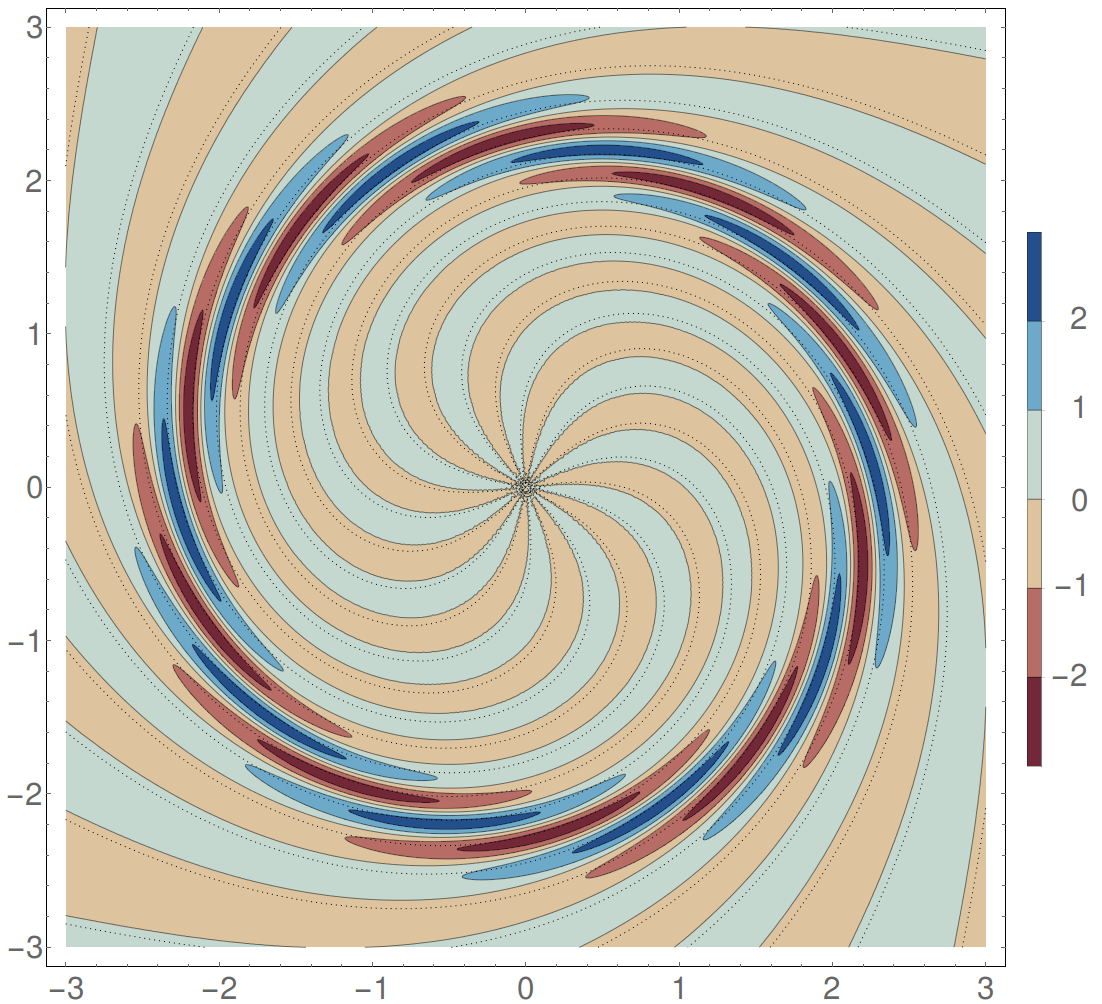}
	\end{center}
	\caption{Function $\chi$ with $l=20, m=10$ and $t/A=2$ in the equatorial plane. Rotation is illustrated by dotted lines which show node lines of function $\chi$ at a later time $t'/A=2.05$.}
	\label{FigPsi20}
\end{figure}
The argument of the cosine function gives the apparent angular velocity of wave shell rotation when the Taylor expansion of the function $\lambda$ is performed at time $t_1$, i.e. $\cos\left( m\varphi-\lambda\left(t,r\right) \right) \approx\cos\left[ m\left(\varphi-\varphi_0-\left(t-t_1\right)\omega_1\right)\right]$, with $\omega _1 = \partial_t \lambda/m$ and $\varphi_0=\lambda\left(t_1,r\right)$. Near the pulse center at $r^2=A^2+t^2$ we get
$\omega_1 = \left(l+1\right)/\left(m A\right)$.  \par
This rotation with $l=m=10$ is illustrated in Figure \ref{FigPsi20} and in the Animation 3 available as an on-line resource \url{http://utf.mff.cuni.cz/~ledvinka/psi/a3.mp4}.
Notice the pulse character of the wave and how the pulse starts and ends as a quite narrow shell. \par
The rotation is related to the angular momentum one can assign to the gravitational waves.
This angular momentum was not considered in \cite{spheres}.
We would now like to discuss, in some detail, the relation betwen the angular momentum and gravomagnetic dragging. \par
To find the angular momentum of the waves we realize that quadratic terms in the second-order Ricci tensor  $R_{\ \ \ jk}^{\left(2\right)}$ play the role of an effective energy-momentum tensor of the waves
as is explained in \cite{spheres} (near (6.2)):
\begin{eqnarray}
	G_{\ \ \ t\varphi }^{(1)}[h^{(2)}] = 8\pi T^{\rm eff}_{{\ \ \ t\varphi }} = - \left\langle R_{\ \ \ t\varphi }^{(2)}\left[h^{(1)} ,h^{(1)}\right]\right \rangle.
\end{eqnarray}
We therefore define the angular momentum as
\begin{eqnarray}
	L_z = - \int T^{\rm eff}_{\ \ \ t\varphi}   \mathrm{d}^3 x
	= \frac{1}{8\pi} \int R^{(2)}_{\ \ \ t\varphi} \mathrm{d}^3 x~.
	\label{Lz}
\end{eqnarray}
This integral can be evaluated using the methods given in \cite{spheres}:
\begin{eqnarray}
\hspace*{-2cm} 
	L_z = {\tilde B}_l^2
	\frac {\left (l+1\right )^{2}\left (l-1\right )\left (l+2\right )^{2}
		lm}
	{\left (5+2\,l\right ){A}^{2}\left (2\,l+3\right 
		)\pi }&
	\left (I^{1/2}_{l+3}\left (2\,{A}^{2}l+3\,{A}^{2}+7\,{t}^{2}+2\,l{t}^{2
	}\right ){A}^{2}\right.\nonumber\\ &\left.+
	I^{3/2}_{l+3}\left (2\,l+3\right )\left ({t}^{2}+{A}^{2}
	\right )^{2}\right ).
\end{eqnarray}
Here the integrals defined in Appendix C of \cite{spheres} give the following results:
\begin{eqnarray}
	I^{1/2}_{l+3}&=
	\frac{\pi}{1+{\tilde t}^2}\frac{\left(2l+3\right)!!}{\left(l+2\right)! 2^{3l+7}},\\
	I^{3/2}_{l+3}&=
	{\pi}\frac{\left(2l+1\right)!!}{\left(l+2\right)! 2^{3l+7}}
	\left[
	\frac{2l+3}{\left(1+{\tilde t}^2\right)^2}+
	\frac{4}{\left(1+{\tilde t}^2\right)^3}
	\right].
\end{eqnarray}
With these integrals, $I^{1/2}_{l+3}$ and $I^{3/2}_{l+3}$ combined, the total value of the angular momentum turns out to be an integral of the motion (assuming $l>1$ to simplify the term $l\left(l-1\right)/l!$) :
\begin{eqnarray}
	L_z = 
	m A^2\,{\tilde B}_l^2\,{\frac {\left (l+1\right ) \left (l+2\right )
			{\left(2l+1\right)!!}}{{2}^{3\,l+6} \left (l-2\right )!}}.
	\label{Lzformula}
\end{eqnarray}
Similar cancellations occur for integral representing energy of linearized  gravitational waves, 
$E =  \int T^{\rm eff}_{\ \ \ tt} \mathrm{d}^3 x$. 
For our packet of gravitational waves an interesting relation holds between both conserved quantities:
\begin{eqnarray}
	L_z = \frac{mA}{l+1}  E.
\end{eqnarray}
\par
The relation between the angular momentum of the gravitational waves and the central frame dragging can be 
seen from the similarity of the integrals \eref{omega0} and \eref{Lz}. Denoting $\mathrm{d}L_z = \left(8\pi\right) ^{-1} R^{(2)}_{\ \ \ t\varphi} r^2 \mathrm{d}\Omega \mathrm{d}r$ and $\mathrm{d}\omega_{(\mathrm{S_w})} = \left(4\pi\right) ^{-1}  R^{(2)}_{\ \ \ t\varphi} r^{-1} \mathrm{d}\Omega \mathrm{d}r$ we get
\begin{eqnarray}
	\mathrm{d}\omega_{(\mathrm{S_w})}  = 2 \frac{\mathrm{d}L_z  }{r^3}.
	\label{Lomegarelation}
\end{eqnarray}
\par
The angular velocity of the central frame dragging can be expressed as an explicit function of the time in the form,
\begin{eqnarray}
	\label{s7eq10}
	\omega_{(\mathrm{S_w})}\left(t\right) = 
	\frac{\tilde B_l^2}{2\pi}\,\frac {m\left (l+1\right )\left (l+2\right )}{A\left (l+3\right )l}
	\times &
	\left[\left( U_l -V_l {\tilde t\,}^2 \right)  \left (1+{\tilde t\,}^{2}\right ) I_{l+3}^2\left(\tilde t\right)\right.\\ &\left.+ \left (U_l+V_l {\tilde t\,}^{2}\right ) I_{l+3}^1\left(\tilde t\right)\right],
\end{eqnarray}
where
\begin{eqnarray}
	\label{s7eq11}
	U_l &= \left (2\,{l}^{5}+7\,{l}^{4}+4\,{l}^{3}-7\,{l}^{2}+24\,l+36\right ),
	\\\nonumber
	V_l &= 3\left ({l}^{4}+2\,{l}^{3}+3\,{l}^{2}+8\,l+12\right )~.
\end{eqnarray}
Here the integrals $I_{l+3}^1$ and $I_{l+3}^2$ are discussed in Appendix C in \cite{spheres}.
Using \eref{Lzformula} we can express $\omega_{(\mathrm{S_w})}$ as a function of the angular momentum 
\begin{eqnarray}
	\omega_{(\mathrm{S_w})}\left(t\right) = \frac{L_z}{2\pi A^3}\,\frac {2^{3l+6}\left(l-2\right)!}{l\left(l+3\right)\left(2l+1\right)!!}
	\times&
	\left[\left( U_l -V_l {\tilde t\,}^2 \right)  \left (1+{\tilde t\,}^{2}\right ) I_{l+3}^2\left(\tilde t\right)\right.\\&\left. + \left (U_l+V_l {\tilde t\,}^{2}\right ) I_{l+3}^1\left(\tilde t\right)\right].
\end{eqnarray}
This result corresponds to the simple relation (52) for dragging by a cylindrical wave.
\par
The dragging becomes maximal at $t=0$ when the wave is closest to the origin:
\begin{eqnarray}
	\omega_{(\mathrm{S_w})}\big|_{t=0} &= \frac{L_z}{4\pi A^3}\,\frac {2^{3l+6}  \left(l+1\right)!\left(l-1\right)!\left(l-2\right)!}{\left(2l+3\right)!\left(2l+1\right)!!}U_l
	\\&=
	\frac{L_z}{A^3}
	\left(
	2+\frac{9}{2l}+\frac {9}{16\,l^2} + ... 
	\right),
\end{eqnarray}
where for $l\ge 4$ these first three terms of expansion provide a relative error less than $ 4\%$.
\par
Finally, neglecting the cosine function in \eref{chi_fce} we find that the maximum of the first order wave occurs approximately at $r=R\left(t\right) \cong \left(A^2+t^2\right)^{\frac{1}{2}}$. 
We then arrive at the simple result demonstrating how the dragging of inertial frame near origin depends on the distance of the (approximate) maximum of the wave and its total angular momentum, it is the same formula as for the material shell:
\begin{eqnarray} \label{fin}
	\omega_{(\mathrm{S_w})}  = \omega_{(\mathrm{S_m})}=\frac{2L_z}{R^3}.
\end{eqnarray}
\section{Conclusions}
In our past work we studied how local inertial frames or gyroscopes are influenced by the motion of both matter and gravitational waves. It is only here, however, where we studied in detail their relationship. By explicit examples  we demonstrated how matter described by standard local energy-momentum tensors drags local inertial frames similarly to gravitational waves satisfying Einstein's equations in vacuum  when a local energy and momentum has not in general been defined.

Since the dragging effects are best illustrated in approximately flat regions surrounded  by rotating matter or waves we considered wave pulses ingoing from infinity, bouncing due their angular momentum and outgoing again to infinity. A suitable analogue made of matter is a collapsing spherical shell made from counter-rotating particles the angular momentum of which can make the shell to rebound like  the waves. If the shell is given a small net rotation, in the first approximation the spacetime inside the shell remains flat but  inertial frames there will rotate with respect to infinity. The analogy cannot be perfect, because the shell rebounds due to its tangential pressure, and this pressure contributes only secondarily to its total angular momentum (by increasing the masss-energy surface density) - the $L_z$ itself comes only from a \textit{perturbative} rotation, whereas it is the $L_z$ of the wave pulse which directly causes it to rebound. Furthermore, unlike the shell, the wave pulse must move at light speed at infinity.

Hence, we first discussed the motion of such a shell.  We derived the motion of the shell from the energy conservation, \eref{eqmotion}.

Next we perturbed the shell by giving it a small net  angular momentum which is conserved during the motion. When compared with the work on slowly rotating shells of dust from particles without counter-rotation, some new effects arise, e.g. for the asymptotic collapse of the shell to and from the region close to the unstable circular orbits of the particles forming the shell. Inside the shell the spacetime remains flat for such small rotations. The dragging of inertial frames inside the shell with respect to infinity can thus be well interpreted.

For the time-symmetric pulse of rotating cylindrical waves ingoing towards the axis of symmetry and then outgoing to infinity again, the rotation of inertial frame at the symmetry axis is, at the moment of time symmetry $t=0$, determined by  the angular momentum per unit length by precisely the same formula as for the cylinder made of dust (cf. expressions (1) and (44)); the same is true at large times when pulse/dust cylinder is near infinity. The values of the dragging are very similar for general positions as well (see \fref{fig:rates}).

In a more complicated case of a time-symmetric ingoing and outgoing regular rotating pulse of gravitational waves in an asymptotically flat spacetime we assumed the waves to have  odd-parity and a non-vanishing angular momentum which keeps them away from the origin where spacetime is very nearly flat, just as it is inside a slowly rotating bouncing material spherical shell. 

When the dragging of the inertial frame due  the pulse near origin is expressed in terms of the distance of the maximum of the pulse and its total angular momentum, we get precisely the same formula as for a slowly rotating spherical shell, \eref{fin}.
\ack
\par
J. B. is a visiting fellow at the Institute of Astronomy in Cambridge. T. L. and J. B. acknowledge 
the partial support from the Grant GA\u CR 17-13525S of the Czech Republic.
W. B. is an undergraduate at the Cavendish Laboratory in Cambridge with a summer studentship at the Institute of Astronomy.
W. B. acknowledges the support of Grant LGOH>EFKM and the hospitality of the Institute of Astronomy.

\appendix
\section{}	\label{A}

The metrics of the exterior and interior are respectively
\be
\mathrm{d}s^{2}=\left(1-r_g/r\right)\mathrm{d}t^{2}-\left(1-r_g/r\right)^{-1}\mathrm{d}r^{2}-r^{2}\left(\mathrm{d}\theta^{2}+\sin^{2}\theta\mathrm{d}\varphi^{2}\right)
\ee
and
\be
\mathrm{d}s^{2}=\mathrm{d}\bar{t}^{2}-\mathrm{d}\bar{r}^{2}-\bar{r}^{2}\left(\mathrm{d}\theta^{2}+\sin^{2}\theta\mathrm{d}\varphi^{2}\right),
\ee
due to spherical symmetry there are common angular coordinates $\theta, \phi$ in both.
The shell constitutes a timelike hypersurface, whose normals in the exterior and interior regions are spacelike, it has the metric,
\begin{eqnarray}
\mathrm{d}s^{2}=\mathrm{d}\tau^{2}-R^{2}\left(\mathrm{d}\theta^{2}+\sin^{2}\theta\mathrm{d}\varphi^{2}\right).
\end{eqnarray}
Defining
\begin{eqnarray} \label{der}
\frac{\mathrm{d}\bar{t}}{\mathrm{d}\tau}\bigg|_{(\mathrm{shell})}=\dot{\bar{T}}=\sqrt{1+\dot{R}^{2}}, \quad \frac{\mathrm{d}t}{\mathrm{d}\tau}\bigg|_{(\mathrm{shell})}=\dot{T}=\frac{\sqrt{1-r_{g}/R+\dot{R}^{2}}}{1-r_{g}/R},
\end{eqnarray}
the four-normals of the hypersurface, each directed \textit{into} their respective spaces in the style of \cite{normals}, are:
\begin{eqnarray}
\left[n_{i}\right]=\left(-\dot{R}, \dot{T}, 0, 0\right), \quad \left[\bar{n}_{i}\right]=\left(\dot{R}, -\dot{\bar{T}}, 0, 0\right).
\end{eqnarray}
An observer whose proper time is $\tau$ has four-velocities $u^{i}$ and $\bar{u}^{i}$. Einstein's equations relate the contents of thin shells to the geometry of their embedding by
\begin{eqnarray} \label{eq:old}
n_{i}\frac{D u^{i}}{D \tau}+\bar{n}_{i}\frac{D \bar{u}^{i}}{D \tau}=8\pi \left(\tau_{\alpha\beta}u^{\alpha}u^{\beta}-\frac{1}{2}\tau^\alpha_\alpha\right),
\end{eqnarray}
where $\tau^{\alpha}_{\beta}$ is the stress-energy tensor defined in the shell:
\begin{eqnarray}
\tau_{\alpha\beta}=\left(\sigma+\Pi\right)u_{\alpha}u_{\beta}
-\Pi \; g_{\alpha\beta}.
\end{eqnarray}
The mass-energy density is $\tau^{0}_{0}=\sigma$, specifically
\begin{eqnarray} \label{eq:equation of state}
\sigma = M/4\pi R^{2}\sqrt{1-\dot{x}^{2}}, \quad \mathrm{d}x^{2}=R^{2}\left(\mathrm{d}\theta^{2}+\sin^{2}\theta\mathrm{d}\varphi^{2}\right)
\end{eqnarray}
which can be re-written in terms of $h$ using the \eref{der},
\begin{eqnarray} \label{eq:angmomentum}
h&=R\left({\mathrm{d}x}/{\mathrm{d}s}\right) 
 =R\left({\mathrm{d}x}/{\mathrm{d}\bar{t}}\right)\left[1-\left({\mathrm{d}R}/{\mathrm{d}\bar{t}}\right)^{2}-\left({\mathrm{d}x}/{\mathrm{d}\bar{t}}\right)^2\right]^{-1/2}
\\ & =R(\mathrm{d}x/\mathrm{d}t)\left[1-r_{g}/R-\left(\mathrm{d}R/\mathrm{d}t\right)^{2}/(1-r_{g}/R)-\left(\mathrm{d}x/\mathrm{d}t\right)^2\right]^{-1/2} \\ & =R\dot{x}/\sqrt{1-\dot{x}^{2}},
\end{eqnarray}
so that $\dot{x}=h/\sqrt{R^{2}+h^{2}}$. The pressure is then $\Pi= M\dot{x}^{2}/8\pi R^{2}\sqrt{1-\dot{x}^{2}}$.
\par
Substituting, the RHS of \eref{eq:old} becomes $-4\pi \left(\sigma+2\Pi\right)$. This equation may be solved as in \cite{shells} by using the orthogonality constraints,
\begin{eqnarray}
u_{i}\frac{D u^{i}}{D \tau}=\bar{u}_{i}\frac{D \bar{u}^{i}}{D \tau}=0,
\end{eqnarray}
to give a second order ODE which may be written as
\begin{eqnarray} \label{eq:transparent}
\hspace*{-2cm} 
\dot{R}\frac{\ddot{R}\left[\sqrt{1+\dot{R}^{2}}-\sqrt{1-r_{g}/R+\dot{R}^{2}}\right]+\frac{m_{g}}{R^{2}}\sqrt{1+\dot{R}^{2}}}{\sqrt{1+\dot{R}^{2}}\sqrt{1-r_{g}/R+\dot{R}^{2}}}
=\frac{m_{g}\dot{R}}{aR^{3}}
\frac{R^{2}+2h^{2}}{\sqrt{R^{2}+h^{2}}}
\end{eqnarray}
and directly integrated to give \eref{eqmotionp}.
The integration constant was not free to choose because Israel's junction conditions have also 
tangential components which yield precisely \eref{eqmotion}. 
If we divide \eref{eqmotion} by $R(t)$,
its left-hand side would be a discontinuity in the $\theta-\theta$ component of the extrinsic curvature of the junction hypersurface, while the right-hand side is equal to $ 8\pi\left(\tau^{2}_{2}- \tau^\alpha_\alpha/2\right)=-4\pi \sigma$. 
Alternatively, \Eref{eqmotionp} may be recovered immediately by substituting our $\sigma$ and $\Pi$ into the result of \cite{normals} which expresses $\sigma$ in terms of desired quantities for more general choices of 
external and internal space.

\section{}	\label{B}
\par
If the shell spins, the projection of the shell hypersurface, $y^\al$, in the exterior space, $x^i$ is corrected by an additional small quantity, $h^{3}_{0}=\dot{\varphi}=\omega_{(\mathrm{S_m})}\dot{T}$,
\begin{eqnarray}
\left[h^{i}_{\alpha}\right]=\left[\frac{\partial x^{i}}{\partial y^{\alpha}}\right]= \left( 
\begin{array}{ccc}
\dot{T} & 0 & 0 \\
\dot{R} & 0 & 0 \\
0 & 1 & 0 \\
\dot{\varphi} & 0 & 1 
\end{array} 
\right),
\end{eqnarray}
but the $n_{i}$ and $\bar{n}_{i}$ are unchanged. It is convenient to keep the definition of the radially attached observer from the perspective of exterior, whose 'proper time' is now an undetermined $\mathrm{d}s^{2}$ rather than $\mathrm{d}\tau^{2}$ as for non-rotating shells. With this in mind,
\begin{eqnarray}
\left[u^{i}\right]=\left(\dot{T}, \dot{R}, 0, 0\right)\frac{\mathrm{d}\tau}{\mathrm{d}s}, \quad \left[\bar{u}^{i}\right]=\left(\dot{\bar{T}}, \dot{R}, 0, \omega_{(\mathrm{S_m})}\dot{T}\right)\frac{\mathrm{d}\tau}{\mathrm{d}s}
\end{eqnarray}
and since $\mathrm{d}s^{2}=\mathrm{d}\tau^{2}-R^{2}\sin^{2}\theta\mathrm{d}\bar{\varphi}^{2}$ within the shell it follows that $\mathrm{d}\tau/\mathrm{d}s$ is unity to ${\mathcal{O}}\left(R\omega_{(\mathrm{S_m})}\right)$. When re-casting \Eref{eq:old} in terms of $s$ rather than $\tau$, the form of the LHS is then unchanged to $\mathcal{O}\left(R\omega_{(\mathrm{S_m})}\right)$ since $\Gamma^{1}_{03}$ is $\mathcal{O}\left(R\omega_{(\mathrm{S_m})}\right)$ and $\Gamma^{1}_{13}$ is identically zero in fully Kerr spacetime. The $\sigma$, $\Pi$  and the inner product of fluid four-velocities in the shell with the $u^{\alpha}$ are also are unchanged to $\mathcal{O}\left(R\omega_{(\mathrm{S_m})}\right)$, accounting for the RHS. From these considerations, the sphericity and equation of motion for the shell must be unchanged to $\mathcal{O}\left(R\omega_{(\mathrm{S_m})}\right)$.
\section{}	\label{C}
\par
The relation \eref{eq:old} stems from the more general result
\begin{equation} \label{eq:einstein}
K_{\alpha\beta}+\bar{K}_{\alpha\beta}=-8\pi \left(\tau_{\alpha\beta}-\frac{1}{2}g_{\alpha\beta}\tau\right),
\end{equation}
where the $K_{\alpha\beta}$ and $\bar{K}_{\alpha\beta}$ are components of the external curvature tensors of the hypersurface in the exterior and interior respectively. Particularly,
\begin{eqnarray}
\tau^{0}_{3}=-\frac{1}{8\pi }K^{0}_{3}=\frac{1}{8\pi }\Bigg(&\dot{T}\left[\frac{\partial \dot{R}}{\partial\varphi}-\Gamma^{0}_{03}\dot{R}+\Gamma^{1}_{03}\dot{T}\right]\nonumber\\&+\dot{R}\left[-\frac{\partial \dot{T}}{\partial\varphi}-\Gamma^{0}_{13}\dot{R}+\Gamma^{1}_{13}\dot{T}\right]\nonumber\\&+\omega_{(\mathrm{S_m})}\dot{T}\left[-\Gamma^{0}_{33}\dot{R}+\Gamma^{1}_{33}\dot{T}\right]\Bigg),
\end{eqnarray}
of which only $\Gamma^{1}_{03}$ and $\Gamma^{0}_{13}$ are nonvanishing to $\mathcal{O}\left(R\omega_{(\mathrm{S_m})}\right)$ and $\Gamma^{1}_{33}$ nonvanishing to $\mathcal{O}\left(1\right)$, giving \eref{rota}:
\begin{eqnarray}
\tau^{0}_{3}=\frac{1}{8\pi }\Bigg(&\dot{T}^{2}\left[R\left(1-\frac{r_{g}}{R}\right)\left(\omega_{(\mathrm{S_m})}+\frac{1}{2}R\frac{\mathrm{d}\omega_{(\mathrm{S_m})}}{\mathrm{d}R}\right)\sin^{2}\theta\right]\nonumber\\&+\dot{R}^{2}\left[-\frac{R^{2}}{2\left(1-\frac{r_{g}}{R}\right)}\frac{\mathrm{d}\omega_{(\mathrm{S_m})}}{\mathrm{d}R}\sin^{2}\theta\right]\nonumber\\&+\omega_{(\mathrm{S_m})}\dot{T}^{2}\left[-R\left(1-\frac{r_{g}}{R}\right)\sin^{2}\theta\right]\Bigg)\nonumber\\&=\frac{1}{16\pi }R^{2}\frac{\mathrm{d}\omega_{(\mathrm{S_m})}}{\mathrm{d}R}\sin^{2}\theta\nonumber\\&=-\frac{1}{16\pi }\frac{\mathrm{d}\omega_{(\mathrm{S_m})}}{\mathrm{d}\frac{1}{R}}\sin^{2}\theta.
\end{eqnarray}

\clearpage
\section*{References}
\bibliography{paper}

\end{document}